# The Thomas Precession Factor in Spin-Orbit Interaction

Herbert Kroemer [*]

Department of Electrical and Computer Engineering,
University of California, Santa Barbara, CA 93106

The origin of the Thomas factor 1/2 in the spin-orbit hamiltonian can be understood by considering the case of a classical electron moving in crossed electric and magnetic fields chosen such that the electric Coulomb force is balanced by the magnetic Lorentz force.

## I. INTRODUCTION: THE PROBLEM

It is well-known that the spin-orbit Hamiltonian differs by a factor 1/2 — the so-called *Thomas Precession factor* — from what one might expect from a naive Lorentz transformation that assumes an uniform *straight-line motion* of the electron. The naive argument runs as follows.

When an electron moves with the velocity **v** through space in the presence of an electric field **E**, the electron will see, in its own moving frame of reference, a Lorentz-transformed magnetic field **B**′ given by the familiar expression

$$\mathbf{B}' = \frac{(\mathbf{E} \times \mathbf{v})/c^2}{\sqrt{1-(v/c)^2}} \xrightarrow{v \ll c} \frac{(\mathbf{E} \times \mathbf{v})}{c^2}, \tag{1}$$

where $c$ is the speed of light; the limit $v \ll c$ is the case of interest here. It is this Lorentz-transformed magnetic field that is supposedly seen by the electron magnetic moment.

This argument suggests that in the presence of electric fields, the magnetic field **B** in the spin Hamiltonian should be replaced by $\mathbf{B} + (\mathbf{E} \times \hat{\mathbf{v}})/c^2$, where $\hat{\mathbf{v}}$ is the velocity operator. This conclusion does not to agree with observed atomic spectra. Historically, this discrepancy provided a major puzzle [1], until it was pointed out by Thomas [2] that the argument overlooks a second relativistic effect that is less widely known, but is of the same order of magnitude: An electric field with a component perpendicular to the electron velocity causes an additional acceleration of the electron perpendicular to its instantaneous velocity, leading to a curved electron trajectory. In essence, the electron moves in a rotating frame of reference, implying an additional precession of the electron, called the *Thomas precession*. A detailed treatment is given, for example, by Jackson [3], where it is shown that this changes the interaction of the moving electron spin with the electric field, and that the net result is that the spin-orbit interaction is cut into half, as if the magnetic field seen by the electron has only one-half the value (1),

$$\mathbf{B}' = \frac{1}{2}\frac{(\mathbf{E}\times\mathbf{v})}{c^2}, \qquad (2)$$

---

[*] Electronic Mail: kroemer@ece.ucsb.edu

It is this modified result that is in agreement with experiment.

A rigorous derivation [3] of this (classical) result requires a knowledge of some aspects of relativistic kinematics which, although not difficult, are unfamiliar to most students. In a course on relativistic quantum mechanics, the result (2) can be derived directly from the Dirac equation, without reference to classical relativistic kinematics. But in a non-relativistic QM course, the instructor is likely to be compelled to an unsatisfactory "it can be shown that" argumentation, which is in fact the approach taken in most textbooks.

In a 1994 textbook of my own [4], I tried to go beyond that approach by considering the case where the electron is forced to move along a straight line, by adding a magnetic field in the rest frame such that the magnetic Lorentz force would exactly balance the electric force. In this case, the Thomas factor of 1/2 was indeed obtained, but the argument — especially its extension to more general cases — lacked rigor. Moreover, by having been published outside the mainstream physics literature, it remained largely unknown to potentially interested readers inside that mainstream. The purpose of the present note is to put the argument onto a more rigorous basis, and to do so in a more readily accessible medium.

## II. CROSSED-FIELD TREATMENT

From our perspective, the central aspect of the "standard" Lorentz transform (1) is the following: The transform $\mathbf{E}_{\text{rest}} \Rightarrow \mathbf{B}_{\text{electron}}$ must be linear in $\mathbf{E}$, and $\mathbf{E}$ can occur only in the combination $\mathbf{E}\times\mathbf{v}$. This combination reflects the two facts that only the component of **E** perpendicular to the velocity **v** can play a role, and that the resulting **B**-field must be perpendicular to both **E** and **v**.

In the absence of any specific arguments to the contrary, we would expect that this proportionality to $\mathbf{E} \times \mathbf{v}$ carries over to the case of a curved trajectory. (For example, a component of **E** parallel to **v** would not contribute to a curved trajectory and to the accompanying rotation of the electron frame of reference.) If one accepts this argument, it follows that $\mathbf{B}' \equiv \mathbf{B}_{\text{electron}}$ should be given by an expression of the general form

$$\mathbf{B}' = \alpha \frac{(\mathbf{E} \times \mathbf{v})}{c^2}, \tag{3}$$

with some scalar proportionality factor $\alpha$ that may still depend on the velocity, but which cannot depend on either **E** or on any magnetic field **B** that may also be present in the rest frame. If there is a magnetic field **B** present in the rest frame, (3) should be generalized to

$$\mathbf{B}' = \alpha \frac{(\mathbf{E} \times \mathbf{v})}{c^2} + \beta \mathbf{B}, \tag{4}$$

where $\beta$ is another constant subject to the same constraints as $\alpha$.

To determine the two constants $\alpha$ and $\beta$, we consider not simply the Lorentz transformation of a pure electric field **E**, but of a specific combination of electric and magnetic fields such that

$$\mathbf{E} = -\mathbf{v} \times \mathbf{B}, \tag{5a}$$

where **v** is the velocity of the electron, and **B** is chosen perpendicular to **v**. For this combination, the electric Coulomb force and the magnetic Lorentz force on the electron cancel, and the electron will move along a straight line. But in this case the simple Lorentz transformation for uniform straight-line motion is rigorously applicable, without having to worry about a rotating frame of reference.

Without loss of generality, we may chose a cartesian coordinate system such that the velocity is in the $x$-direction, the magnetic field in the $z$-direction; and the electric field in the $y$-direction,

$$E_y = v_x B_z. \tag{5b}$$

If this combination of the two fields is Lorentz-transformed into the uniformly moving frame of the electron, the magnetic field $B_z'$ in that frame is

$$B'_z = \frac{B_z - E_y v_x / c^2}{\sqrt{1 - (v_x/c)^2}}. \tag{6}$$

Inasmuch as we are interested only in the limit $v \ll c$, we may expand (6) by powers of $v_x$. The result may be written

$$B'_z = B_z + \frac{B_z v_x^2}{c^2} \cdot \left[\tfrac{1}{2} + \tfrac{3}{8}\left(\frac{v_x}{c}\right)^2\right] - \frac{E_y v_x}{c^2} \cdot \left[1 + \tfrac{1}{2}\left(\frac{v_x}{c}\right)^2\right] + \ldots, \tag{7}$$

where the dots represent omitted terms of order higher than $v_x^4$. But under our condition (5b), we have $B_z v_x^2 = E_y v_x$, and (7) may be simplified by combining terms, to read

$$B'_z = B_z - \frac{E_y v_x}{c^2} \cdot \left[\tfrac{1}{2} + \tfrac{1}{8}\left(\frac{v_x}{c}\right)^2\right] + \ldots \tag{8}$$

Going to the limit $v \ll c$, we obtain finally

$$B'_z = B_z - \tfrac{1}{2}\frac{E_y v_x}{c^2}. \tag{9a}$$

In three-dimensional vector form,

$$\mathbf{B}' = \mathbf{B} + \tfrac{1}{2}\frac{\mathbf{E} \times \mathbf{v}}{c^2}, \tag{9b}$$

which is of the required form (4), with $\alpha = 1/2$ and $\beta = 1$.

This is the magnetic field that is seen by the intrinsic magnetic moment of the electron. It determines the potential energy of that moment in the presence of both electric and magnetic fields for our specific combination of crossed fields, up to orders linear in $\mathbf{E} \times \hat{\mathbf{v}}$. The leading $\mathbf{B}$-term is already included in the non-relativistic spin hamiltonian; the remainder is the first-order correction due to the spin-orbit interaction. Note that it contains the Thomas factor 1/2, in agreement with (2).

The rest is straightforward. Our argument suggests that the magnetic field $\mathbf{B}$ in the spin hamiltonian should be replaced with an operator that is equivalent to (9b). Following standard textbook arguments, we obtain the familiar spin-orbit contribution to the hamiltonian

$$\hat{H}_{\text{SO}} = \mu_e \hat{\boldsymbol{\sigma}} \cdot \hat{\mathbf{B}}' = \frac{\mu_e}{2c^2} \hat{\boldsymbol{\sigma}} \cdot (\mathbf{E} \times \hat{\mathbf{v}}), \tag{10}$$

where $\mu_e$ is the intrinsic magnetic moment of the electron, and $\hat{\mathbf{v}}$ is now the *operator* for the velocity.

## III. DISCUSSION

The central assumption in our derivation was that the proportionality of $\mathbf{B}'$ to $\mathbf{E} \times \hat{\mathbf{v}}$ carries over to the case of a rotating frame of reference. This is of course an exact result. But in a paper explicitly dedicated to the classroom *didactics* of the Thomas factor without invoking less-well-known aspects of relativistic kinematics, it is probably best to treat it as an eminently plausible assumption.

I close with a comment on our retention of the $B_z v_x^2$-term in (7), albeit converted into a $E_y v_x$ -term. This term takes the place of rotating-frame corrections in the case of a pure electric field. Neglecting it would be *exactly* equivalent to neglecting the effects of a rotating frame of reference for a more general choice of fields.